# Prediction of low–Z collinear and noncollinear antiferromagnetic compounds having momentum-dependent spin splitting even without spin-orbit coupling

Lin-Ding Yuan[1], Zhi Wang[1*], Jun-Wei Luo[2], Alex Zunger[1*]

[1]Energy Institute, University of Colorado, Boulder, CO 80309, USA

[2]State Key Laboratory for Superlattices and Microstructures, Institute of Semiconductors, Chinese Academy of Sciences, Beijing 100083, China

*Email: alex.zunger@gmail.com; Zhi.Wang@colorado.edu

Recent study (Yuan et. al., Phys. Rev. B 102, 014422 (2020)) revealed a SOC-independent spin splitting and spin polarization effect induced by antiferromagnetic ordering which do not necessarily require breaking of inversion symmetry or the presence of SOC, hence can exist even in centrosymmetric, low-Z light element compounds, considerably broadening the material base for spin polarization. In the present work we develop the magnetic symmetry conditions enabling such effect, dividing the 1651 magnetic space groups into 7 different spin splitting prototypes (SST-1 to SST-7). We use the 'Inverse Design' approach of first formulating the target property (here, spin splitting in low-Z compounds not restricted to low symmetry structures), then derive the enabling physical design principles to search realizable compounds that satisfy these a priori design principles. This process uncovers 422 magnetic space groups (160 centrosymmetric and 262 non-centrosymmetric) that could hold AFM-induced, SOC-independent spin splitting and spin polarization. We then search for stable compounds following such enabling symmetries. We investigate the electronic and spin structures of some selected prototype compounds by density functional theory (DFT) and find spin textures that are different than the traditional Rashba-Dresselhaus patterns. We provide the DFT results for all antiferromagnetic spin splitting prototypes (SST-1 to SST-4) and concentrate on revealing of the AFM-induced spin splitting prototype (SST-4). The symmetry design principles along with their transformation into an Inverse Design material search approach and DFT verification could open the way to their experimental examination.

## I. Introduction

Manipulating the spin as well as spin polarized electrons in solids is a central interest of spintronics [1-3]. The conventional way of creating and manipulating spin polarization and magnetization entails use of the (a) Rashba-Dresselhaus effects[4,5] involving spin-orbit coupling (SOC) induced spin splitting in non-centrosymmetric, non-magnetic, heavy-element materials susceptible to external electric fields[6], and (b) the Zeeman effect[7,8] in ferromagnets (FM) having non-zero net magnetization when external magnetic fields are applied. The SOC-induced spin splitting and consequent spin polarization have been known to generate spin-orbit torque[9], while the ferromagnetic spin polarization has been widely used for spin generation and detection [1]. On the other hand, antiferromagnets (AFM) have alternate local magnetic moments on different atomic sites, which mutually compensate to a global zero net magnetization, making AFM immune to the effect of external magnetic fields. A recent theoretical study [10] pointed out that spin splitting of energy bands and spin polarized electron states could be present in antiferromagnets possessing specific spatial and magnetic symmetries. Unlike the Zeeman effect in ferromagnets, such AFM-induced spin splitting and spin polarization arise from coupling of the position coordinate $r$ of the intrinsic inhomogeneous magnetic field $h(r)$ set up, e.g. by superexchange interactions in AFM with the electron spin $\sigma$. Significantly, large magnitude of spin splitting comparable to the best known ('giant') SOC effects has been illustrated theoretically via realistic density functional theory (DFT) calculations in as rutile MnF$_2$[10]. It has been shown that unlike the Rashba-Dresselhaus effect, such AFM effect is present even in centrosymmetric crystals and without relativistic spin-orbit coupling (SOC), hence could eliminate the challenges of weak chemical bonds [11] and undesirable defects often present in high Z compounds having large SOC such as rare-earth compounds as well as Tellurides and Antimonides [12-14]. This AFM magnetic mechanism of spin splitting could enable AFM to assume an active role in spintronics. In this work we use the Inverse Design approach [15-19] of first formulating the target property for spin splitting then deriving the causal design principles (here, the spatial and magnetic symmetry conditions) to search realizable compounds that satisfy these *a priori* design principles, and finally illustrating examples by DFT calculations. The main steps after establishing the target property of spin splitting in low-Z compounds not restricted to low symmetry structures, are:

(i) We formulated the spatial and magnetic symmetry design principles (DP's) that enable the different prototypes of spin splitting and polarization effects. In doing so, we place the physics of AFM spin splitting within the broader context of symmetry conditions that enable the more familiar forms of spin splitting such as (a) ferromagnetic Zeeman effect [7,8], the (b) Rashba and Dresselhaus effects [4,5],the (c) ordinary centrosymmetric AFM compounds that have no spin splitting, and (d) those that have SOC-induced spin splitting that can exist within background antiferromagnetism [20]. Special attention is given to the symmetry conditions that enable pure AFM-induced spin splitting that could exist even in the absence of SOC (non-relativistic Hamiltonian) and even in centrosymmetric structures -- the so-called AFM-induced spin splitting prototype 4A and 4B both collinear and noncollinear.

(ii) Based on these DP's, we developed the formal procedures for determining the spin splitting prototype (SST) starting from identification of magnetic space group. We use this approach to identify crystallographic and magnetic compound databases for examples that satisfy each of the seven SST.

(iii) We examined the band structures and spin textures of specific subset of these including both collinear and noncollinear AFM. Like SOC, noncollinear AFM can creates spin polarization that varies in direction in momentum space.

This work then provides the foundation of spin polarization materials, offering also a bridge between such design principles and recognizable crystal and magnetic structures.

## II. SYMMETRY PRINCIPLES FOR IDENTIFYING DIFFERENT SPIN SPLITTING PROTOTYPES

### A. Essential features and classification of magnetic space groups

Except for accidental degeneracy, the degeneracy of spin states is ensured by specific symmetries. In a non-magnetic compound, the breaking of the spatial inversion symmetry $I$ is necessary for (e.g., Rashba and Dresselhaus[4,5]) spin splitting. In contrast, in a magnetic compound with zero net magnetization (e.g. no Zeeman effect[7]), understanding the violation of spin degeneracy requires analysis of the magnetic space group (MSG) symmetry[21].

A magnetic space group is a symmetry group that consists of all symmetry operators that retain the invariance of both the atomic structure and the magnetic order. Different from the traditional space group, the magnetic space groups (denoted for brevity in equations as $M$) are composed of both unitary space symmetries $G = \{R_i\}$ and anti-unitary symmetries $A = \{\theta R_m\}$, that is $M = G + A$. The unitary part $G$ consists of pure spatial operations (rotations, translations, and/or their combinations), while the antiunitary part $A$ consists of combined symmetries of time reversal and spatial operations. A parent space group (PSG) consists of spatial symmetries that keep the atomic structure invariant. It not only includes the unitary set of symmetries $G$ of the magnetic space group, but also other spatial symmetries that keep the atomic structure invariant but change the magnetic order.

The magnetic space group symmetries of a compound are determined not only by the positions of its magnetic ions but also by the positions of the non-magnetic ions in the unit cell. Indeed, theoretical simplifications neglecting the nonmagnetic (ligand) ions in the lattices might lead to incorrect predictions of spin splitting behavior. For example, the AFM-induced spin splitting evident in DFT calculations in tetragonal $MnF_2$ in the absence of SOC [10] will vanish if considering a model retaining only the magnetic sublattice $Mn^{2+}$ and neglecting the $F^-$ ligands. Such decisive effect of non-magnetic atoms on the spin splitting behavior highlights the possible important role of the (super)exchange interaction mediated via the non-magnetic atomic sublattice.

The magnetic space groups are classified into four MSG types [21] in terms of the properties of the unitary symmetry set $G$ and anti-unitary symmetry set $A$. MSG Type I ('colorless') has only unitary symmetries, i.e. $A = \emptyset$ hence $M = G$. There are in total 230 MSG's that belong to MSG Type I, the same number as of the space group. MSG Type II is the 'grey group' with $A = \theta G$ hence $M = G + \theta G$, i.e., for each unitary symmetry $R_i$ there is a correspondence nonunitary symmetry $\theta R_i$ in the MSG. Non-magnetic compounds under zero external magnetic field belong to this MSG. There are also 230 MSG's belonging to MSG Type II. MSG Type III (517 in total) and Type IV (674 in total) are known as 'black-white' groups with $M = G + aG$ (where $a$ is an antiunitary symmetry of $M$). The unitary part $G$ is then the invariant subgroup of $M$ of index 2 ($A = aG$, then $G$ and $A$ have equal number of elements). The distinguishing feature of MSG type III compound from MSG type IV compound is that the latter contains a translation $T$ that reverses the direction of the magnetic order, therefore has symmetry $\theta T$ (refereed as anti-translation symmetry), while MSG type III does not have $\theta T$ symmetry. An alternative way to distinguish whether a compound belongs to MSG type III or IV is via the relation between its magnetic and non-magnetic unit cells: If a compound has a magnetic primitive unit cell that is equivalent to a supercell of its non-magnetic primitive unit cell, then it has $\theta T$ symmetry, and consequently is MSG type IV. On the other hand, if a compound has a magnetic primitive unit cell equivalent *to* the non-magnetic primitive unit cell itself (not to the supercell), then it has no $\theta T$ symmetry, hence cannot belong to MSG type IV (it can belong to MSG type III, or MSG type I). Generally, AFM compounds can be MSG type I, type III or type IV, depending on whether the antiunitary set $A$ is empty, not empty but has no $\theta T$, or has $\theta T$, respectively.

B. **Magnetic symmetry requirements for the seven spin splitting prototypes**

1. **Design principles**

There are two symmetry design principles (DP's) for the presence of spin splitting and spin polarization:

*(i) Finite spin splitting requires violation of ΘI and ΘIT symmetries (referred to as "DP-I")*. The combination $\theta I$ of time reversal $\theta$ and spatial inversion $I$ ensures double spin degeneracy for arbitrary wave vector $\boldsymbol{k}$, providing the transformation of energy dispersion under time reversal is $\theta E(\boldsymbol{k},\boldsymbol{\sigma}) = E(-\boldsymbol{k},-\boldsymbol{\sigma})$ and under inversion is $IE(\boldsymbol{k},\boldsymbol{\sigma}) = E(-\boldsymbol{k},\boldsymbol{\sigma})$. In addition to this, the combination of $\theta I$ with an additional translation $T$, that leads to $TE(\boldsymbol{k},\boldsymbol{\sigma}) = E(\boldsymbol{k},\boldsymbol{\sigma})$, will also preserve double spin degeneracy. That means if the system has $\theta I$ or $\theta IT$ symmetry the bands will be degenerate at any k point. Therefore, the appearance of spin splitting requires the violation of $\theta I$ and $\theta IT$ symmetries.

*(ii) The presence of spin splitting when SOC is absent requires MSG type I or III (referred to as "DP-II")*. When and only when SOC is turned off ( i.e. non relativistic Hamiltonian) , there could exist S=1/2 spinor symmetry $U$, which reverses the spin and direction of magnetic order but retains the wavevector invariance i.e., $UE(\boldsymbol{k},\boldsymbol{\sigma}) = E(\boldsymbol{k},-\boldsymbol{\sigma})$. The presence of $U$ symmetry then preserves spin degeneracy for all wavevectors. Such $U$ symmetry is present in all *non*-magnetic compounds of MSG type II, and accounts for spin degeneracy when SOC is absent. In contrast, in magnetic compounds where non-zero magnetic moments are located on different atomic sites, $U$ cannot be a symmetry operation since $U$ reverses the direction of magnetic order. In MSG type IV antiferromagnetic compounds, where a translation $T$ that reverses the direction of magnetic order is present, there is $UT$ symmetry (combination of $U$ with translation $T$) that would also preserve spin degeneracy for all wavevector. As opposed to AFM of MSG type IV where sublattices of opposite spin are related by $UT$ symmetry, AFM of MSG type I or III does not hold such UT symmetry. For another viewpoint, AFM of MSG type I or III, where the magnetic unit cell is identical to the nonmagnetic unit cell, has a sublattice degree of freedom with opposite spins on alternating sublattices, and spin splitting can be created without requiring the Zeeman and SOC mechanism. As a result, of the four MSG types, the appearance of spin splitting when SOC is off requires MSG type being I or III (has no $UT$ symmetry).

2. **Definitions of seven spin splitting prototypes**

Depending on which of the two design principles DP-I and DP-II for spin splitting are satisfied, we identified seven spin splitting prototypes including four AFM spin splitting prototypes, one FM prototype, and two NM prototypes (Figure 1):

**SST-1** (*AFM without spin splitting)* are AFM compounds violating DP-I but satisfying DP-II. These are centrosymmetric AFM compounds that preserve the ΘI (or ΘIT) symmetry and have type III magnetic space group i.e. have a magnetic primitive unit cell *equivalent to* the non-magnetic primitive unit cell. The existing ΘI (or ΘIT) symmetry (violation of DP-I) then ensures no spin splitting for both SOC off and SOC on cases. Example of SST-1 include tetragonal AFM CuMnAs[22] with magnetic space group Pm'mn.

**SST-2** (*AFM without spin splitting*) are AFM compounds violating both DP-I and DP-II. These are centrosymmetric AFM compounds that preserve the ΘI (or ΘIT) symmetry and (in contrast with SST-1) have type IV magnetic space group i.e. the magnetic primitive unit cell is a supercell of the non-magnetic primitive unit cell. SST-2 differs from SST-1 in MSG type, while the present of ΘI (or ΘIT) symmetry again results in no spin splitting for both SOC off and SOC on cases. Example of SST-2 include rocksalt AFM NiO[23] with magnetic space group $C_c2/c$.

| Magnetic Structure | Spin splitting prototype (SST) | Main feature | Symmetry conditions | | | Consequences of symmetry conditions on SS at generic $k$ | |
|---|---|---|---|---|---|---|---|
| | | | Is $\theta I$ or $\theta IT$ in MSG? | MSG type | Is the crystal CS or NCS? | when SOC is off | when SOC is on |
| AFM | SST-1 | AFM, No spin splitting | Yes | III | CS | No | No |
| | SST-2 | AFM, No spin splitting | Yes | IV | CS | No | No |
| | SST-3A | SOC-induced (AFM as background) | No | IV | CS | No | Yes |
| | SST-3B | | No | IV | NCS | | |
| | SST-4A | AFM-induced | No | I/III | CS | Yes | Yes |
| | SST-4B | | No | I/III | NCS | | |
| FM | SST-5 | Zeeman effect | No | I/III | Either | Yes | Yes |
| NM | SST-6 | NM, No spin splitting | Yes | II | CS | No | No |
| | SST-7 | Rashba and Dresselhaus effect | No | II | NCS | No | Yes |

**Figure 1 | Classification of spin splitting prototypes (SSTs) in terms of symmetry conditions and the consequences of these symmetry conditions on the spin splitting with or without SOC.** Here, $I$, $\theta$ and $T$ are symmetry operations of spatial inversion, time reversal, and translation, respectively. $\theta I$ and $\theta IT$ are combinations of these operators. The symmetry conditions are phrased in terms of three questions: (1) Is either the symmetry $\theta I$ or $\theta IT$ present in the magnetic space group (MSG)? (DP-I) (2) What is the MSG type? (DP-II) (3) Is the parent space group (PSG) centrosymmetric (CS) or not (NCS)? The consequences of symmetry on the spin splitting patterns at a generic k point are given for both cases with and without SOC.

**SST-3** (*SOC- induced spin splitting in the presence of AFM*) are AFM compounds satisfying DP-I but violating DP-II. These compounds violate the ΘI (or ΘIT) symmetry and have type IV magnetic space group. Although they are magnetic, the underlining net magnetization is zero, and these compounds behave similarly to non-magnetic and non-centrosymmetric conventional Rashba[4] and Dresselhaus[5] compounds (described later as SST-7), creating spin splitting only when SOC is turned on, and the magnitude of spin splitting is proportional to the strength of SOC. SST-3 can thus be viewed as a special case of traditional SOC-induced spin splitting where the existence of 'background AFM' does not interfere with the spin splitting but does not create it in its own right. Unlike nonmagnetic compounds, the violation of ΘI (or ΘIT) symmetry in the AFM compounds does not mean the violation of inversion symmetry. The crystal of the AFM compounds can be either centrosymmetric or non-centrosymmetric. A classification of centrosymmetric vs non-centrosymmetric prototypes then helps identify the interesting cases of centrosymmetric AFM crystals having spin splitting, contrasting them with the traditional Rashba and Dresselhaus cases of non-centrosymmetric spin splitting. SST-3 is then divide into two subprototypes: SST-3A being centrosymmetric (e.g. $MnS_2$) and SST-3B is non-centrosymmetric (e.g. $BiCoO_3$). Example of the SST-3B is $BiCoO_3$[20] that has been shown to have vanishing spin splitting when SOC is turned off and non-zero spin splitting when SOC is turned on.

**SST-4** (*AFM-induced spin splitting even without SOC*) are AFM compounds that satisfy both the DP-I and the DP-II. These are AFM compounds violating the ΘI (or ΘIT) symmetry and have magnetic space group type I or III. In a way, SST-4 prototypes are the most interesting cases where *spin splitting is present in the absence of SOC and under zero net magnetization*. That spin splitting survives in SST-4 even without SOC term in the Hamiltonian[10] while maintaining zero net magnetization implies that it is induced by mechanisms other than SOC and Zeeman. Interestingly, such AFM-induced spin splitting can be very large even in low atomic number compounds where SOC is negligible, and consequently does not rely on the often-unstable high-Z elements required for large SOC. When SOC is turned on in the Hamiltonian of SST-4 on top of the AFM- induced spin splitting, new spin splitting can emerge even at time reversal invariant moments (TRIMs). Similar to SST-3, the crystal of SST-4 compounds can be either centrosymmetric or non-centrosymmetric. A classification of centrosymmetric vs non-centrosymmetric prototypes then divides SST-4 into two subprototypes: SST-4A being centrosymmetric (e.g. Pnma $LaMnO_3$) and SST-4B is non-centrosymmetric (e.g. R3c $MnTiO_3$).

**SST-5 (***Zeeman spin splitting in ferromagnets***)** are FM compounds satisfying both the DP-I and the DP-II. These are ferromagnetic compounds that violate $\theta I$ (or $\theta IT$) symmetries and have magnetic space group of type I or III. Just like AFM SST-4, ferromagnets SST-5 have spin splitting for SOC off and on. But the resulting spin splitting in FM is induced by the nonzero net magnetization. The underlining Zeeman mechanism is SOC unrelated and gives rise to spin splitting even when SOC is turned off. All ferromagnets, or more generally any magnetic compound that has non-zero net magnetization (such as ferromagnetic Fe) belong to SST-5.

**SST-6** (*nonmagnetic with no spin splitting*) are nonmagnetic compounds that violate both the DP-I and the DP-II. These are compounds preserving $\theta I$ (or $\theta IT$) symmetry. Just as the centrosymmetric AFM SST-1 and SST-2, the presence of $\theta I$ (or $\theta IT$) in SST-6 compounds then guarantees zero spin splitting. Well known example of SST-6 includes centrosymmetric non-magnetic semiconductor Si and Ge.

**SST-7** (*SOC-induced nonmagnetic: Rashba-Dresselhaus*) are the traditional nonmagnetic Rashba-Dresselhaus compounds satisfying DP-I but violating DP-II. These compounds violate $\theta I$ (or $\theta IT$) symmetry. The violation of $\theta I$ (or $\theta IT$) in non-magnet where time reversal symmetry $\theta$ is present, is equivalent to violation of inversion symmetry $I$, which gives rise to spin splitting effects, known as Rashba[24] and Dresselhaus[5] effect. Non-centrosymmetric non-magnetic semiconductor GaAs belongs to this category.

3. **Grouping of Spin splitting prototypes according to the consequences of the symmetry conditions**

The seven SSTs can be grouped into three fundamental types of consequences (Fig. 1):

(a) *SSTs that have no spin splitting either with or without SOC*: SST-1, SST-2, and SST-6. The zero spin splitting happens at every k point, and associates with vanishing global spin polarization. As been pointed out by Zhang et. al. [25], in a centrosymmetric nonmagnetic crystal (SST-6) where spin splitting and global spin polarization are forbidden by symmetry, local spin polarization would still present, this is known as "hidden spin polarization" [25]. Resembling to the "hidden spin polarization" in SST-6, globally vanished but locally present spin polarization effect would also exist in AFM prototype SST-1, SST-2 where no spin splitting presents.

(b) *SSTs that have spin splitting only when SOC is present*: SST-3A, SST-3B, and SST-7. In these cases, the spin splitting is purely SOC induced. The corresponding spin polarization varied at different k points and would give non-trivial helical (e.g. Rashba-Dresselhaus[4,5]) spin texture.

(c) *SSTs that have spin splitting either with or without SOC*: SST-4A, SST-4B, and SST-5. In these cases, the spin splitting could be SOC unrelated and the spin states are polarized mainly in the direction of the magnetization. For collinear magnets, the spin is collinearly polarized and conserved; for noncollinear magnets, the spin is expected to be noncollinearly polarized and resembles to the SOC induced momentum-dependent spin polarization.

4. **Collinear spin polarization in collinear AFM compounds and noncollinear spin polarization in noncollinear AFM compounds:**

An important aspect of AFM-induced spin splitting depends on whether the spins are arranged collinearly or noncollinearly. The symmetry-based classification of SSTs in section II is generally applicable not only for collinear magnetization, but also for non-collinear and incommensurate magnetizations. For example, one would expect AFM-induced spin splitting in non-collinear centrosymmetric AFM $Mn_3Ir$ [26] with MSG R-3m' (MSG type III, no ΘI or ΘIT symmetry). Therefore, both collinear and non-collinear AFM compounds can hold nonzero spin splitting and spin polarization even in the absence of SOC.

In *collinear AFM*, the spin is collinearly polarized and conserved. The spin polarization collinearity disallows the system to generate dissipationless charge or spin current (such as spin Hall effect, or anomalous Hall effect) by electric field [27,28] or give rise to current-driven magnetization on its own in the nonrelativistic limit. Such limitations of collinear spin polarization in collinear AFM can be overcome by other mechanism within spin collinearity that allows similar applications. Specifically, despite the absence of spin Hall effect or anomalous Hall effect, collinear AFM allows *magnetic spin Hall effect* and longitudinal spin polarized flowing electrons, as shown in collinear AFM compound $RuO_2$[57]. Moreover, although the collinear spin polarization cannot give rise to current induced torque as would induced by SOC, it would instead be used to generate Spin Transfer Torque (STT) in a magnetic tunnel junction or spin valve which might drive magnetization switching in other materials (not in the material itself).

In *noncollinear AFM*, the spin is noncollinearly polarized. Such noncollinear spin polarization can induce effects that resemble SOC-related effects even without the presence of SOC, such as the anomalous Hall effect[55] and spin hall effect[56]. However, similar to SOC-related effect, because of the noncollinearity, spin is not conserved in momentum space. The momentum dependent spin precession together with the momentum scattering causes spin dephasing and shorter spin lifetime.[29]

On one hand, the collinear magnets preserve spin and would enable long spin lifetime; on the other hand, the noncollinear magnets would generate noncollinear spin polarization in momentum space which resembles the relativistic SOC effects and gives rise to many exotic physical phenomena like anomalous Hall effect[55] and spin Hall effect[56] etc. But unlike SOC induced spin polarization the noncollinear spin texture induced by noncollinear magnetization is unique to the magnetization and reflects the symmetry of the magnetic structure.

5. **Association of the 1651 magnetic space groups into seven spin splitting prototypes**

Given the magnetic space group, one can determine whether it has ΘI or ΘIT symmetry or not and what is the magnetic space group type. It is then straightforward to determine the spin splitting prototype and predict the spin splitting consequences of one compound based on the violation or satisfaction of the two design principles given above. **Table AI** in Appendix I lists all the 1651 three-dimensional magnetic space groups that are classified into the seven spin splitting prototypes. For SST-3 and SST-4, additional information of the corresponding parent space group might be required in case to determine its sub-prototype A or B. For magnetic space groups that do not include $\theta I$ symmetry, as listed in the fifth row of Table I, both DP-I and DP-II are satisfied therefore it allows spin splitting even when SOC is absent. Compounds with magnetic space group in this category are either SST-4 or SST-5, depending on their magnetic type being antiferromagnet or ferromagnet.

In summary, Section II derived two symmetry design principles for the occurrence of spin splitting even without spin-orbit coupling. Based on the two design principles and the magnetic types we have defined seven different spin splitting prototypes. The classification of different spin splitting prototypes would then guide the searching of materials.

## III. From formal definitions of seven spin splitting prototypes to the identification of compounds that belong to them

Whereas in the foregoing section and Table AI we sorted magnetic space groups (but not magnetic compounds) into the seven spin splitting prototypes, in the present section, following the Inverse Design paradigm, we outline the steps needed to sort specific compounds into the functionality of seven spin splitting prototypes. The steps are:

(i) *Given the crystal structure and magnetic moment configuration of a compound, find its magnetic space group.* Starting from the crystal structure and magnetic moment arrangement one can determine the magnetic symmetries by explicitly examine the invariance of the magnetic structure under spatial symmetries of its parent space group and their corresponding combination with time reversal symmetry. There are a few tools helping with the identification, such as "FINDSYM" developed by Stokes et.al.[30] It identifies the magnetic space group of a compound given the position of the atoms and magnetic moments arrangements in the unit cell.

(ii) *Associate a given compound with a given SST.* For a given magnetic compound, the formal procedure of determining its SST after knowing the magnetic space group is to check from Appendix Table AI which spin splitting prototype the magnetic space group belongs to. If it belongs to the fourth category, one will need to check its magnetic type. If it is an AFM, then it belongs to SST-4; if it is an FM, then it belongs to SST-5.

The identification of spin splitting behavior via direct inspection of its atomic and magnetic structure is also possible by checking: (1) Whether there is inversion center in the crystal that interchanges the magnetic moments of opposite orientation; (2) whether there is pure translation that interchanges the magnetic moments of opposite orientation. An answer of (yes, yes) corresponds to violation of DP-I and violation of DP-II. Moreover, propagation vector of a magnetic compound determined in a neutron diffraction can also provide useful information about the SST. Magnetic compound with at least one propagation vector component being 1 or a fraction of even denominator (e.g., (1,0,0), (1/2,0,0) etc.), usually belongs to MSG-type IV, i.e., violate DP-II. **Table AII** in Appendix I gives our symmetry based classification of magnetic compounds currently listed in Bilbao MAGNDATA database of commensurate magnetic structures [31] as AFM SST-4A and 4B.

(iii) *Down selecte the best compounds from the list deduced from symmetry according to practical considerations.* Many of the symmetry defined candidates shown in Table AII are not in their ground state, or might have unwanted toxic elements, or might be stable only in alloy form. To select compounds that are more useful for future application, we (1) choose stable and simple magnetic structures; (2) we remove alloys from the list; and (3) we select preferably low-Z compounds. **Table AIII** in Appendix I gives the classification of compounds currently listed in Bilbao MAGNDATA[31] according to the presently defined SST-1 to SST-4 (Fig.1), as well as collinear and noncollinear, with additional filters applied (1) experimentally synthesized; (2) not a disordered alloy; (3) composed of low-Z, non-toxic elements. These compounds are the most interesting cases since the spin splitting and spin polarization are inherit to its AFM ordering and do not require SOC.

(iv) *Compare DFT calculation of the spin splitting profile of given compound with the expectation based on its classification as SST in Fig.1*. Given the position of the atoms and magnetic moments arrangements in the unit cell of item (ii) one can independently calculate its band structure and extract spin splitting and spin texture. Such predictions are then associated with the *a priori* predictions of Fig.1 based on the classification of the said compound into SST.

In summary, section III provides the formal procedures for determining SST for one compound starting from knowing its magnetic space group. We also discussed empirical method for directly inspecting the SST. We applied the method to many known magnetic compounds and identified a list of AFM SST-4 compounds. We further down selected these compounds into a smaller set of candidates consisting of low-Z, non-toxic elements as favored by real applications.

# IV. Density Functional illustrations of the electronic structure, spin splitting of representative examples of antiferromagnetic spin splitting prototypes

The electronic structures are calculated by the DFT [32] using a plane wave basis set and the exchange correlation functional of Perdew-Burke-Ernzerhof (PBE)[33,34] with on-site Coulomb energy accounted by an effective U parameter[35] following the simplified rotationally invariant approach introduced by Dudarev et. al.[36]. Experimental crystal and magnetic structure are used for the DFT modeling. A Γ-centered k-mesh is used for hexagonal cells and Monkhorst-Pack[37] k-mesh for other crystals. The key features of spin splitting and spin polarization are examined within the same DFT frame. In the calculations, the spin splitting is evaluated for the difference in the eigen values of neighboring bands holding opposite spin polarization, while the spin polarization for Bloch state $|\mathbf{k}\rangle$ at momentum $\mathbf{k}$ is calculated as the expectation of the spin operator.

To reveal the explicit electronic and spin properties, we have examined representative compounds for each AFM spin splitting prototype from SST-1 to SST-4. DFT results for the most interesting cases of SST-4A and SST-4B for both collinear and noncollinear compounds are discussed in detail.

## A. Tetragonal, P4/nmm, CuMnAs illustrating no spin splitting SST-1

We use the experimental crystal structure from Ref. [22] for the tetragonal CuMnAs and set the effective *U* on Mn 3d orbits to 1.9 eV as the input for DFT electronic band structure calculations. Figure 2(a) shows the crystal structure of the AFM phase of the tetragonal CuMnAs which can be stabled on GaP surface. We apply magnetic moment on Mn initially along the [001] direction, and obtain from the calculation a final magnetic moment on Mn of 4.1 $\mu_B$ aligned along the [001] direction (see AFM magnetization contour on selected plane in Fig. 2(b)) which agrees with the neutron-diffraction measurement of 3.6 $\mu_B$[22]. Figure 2(c) shows the first Brillouin zone and high symmetry k-paths. Figure 2(d) and (e) show the corresponding spin degenerate bands with SOC off (d) and on (e). The bands show metallic behavior with no band gap near Fermi level for both SOC off and on. Small difference on band structure near Fermi level has been predicted between (d) and (e), which is expected as all elements are light elements. Such spin degenerate bands in the whole Brillouin zone even when SOC is on, are in agreement with our prediction of no spin splitting for SST-1.

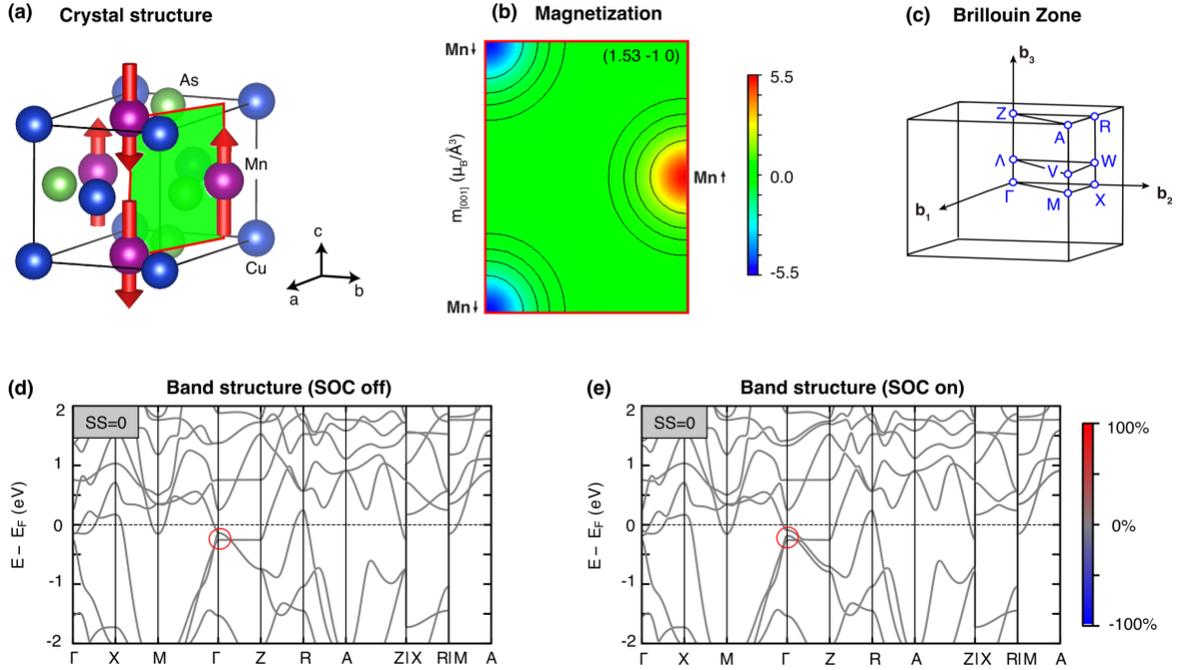

**Figure 2 | No spin splitting in AFM CS tetragonal CuMnAs (AFM SST-1).** (a) Crystal structure and magnetic moments, where red arrows are used to indicate local magnetic moments; (b) z-component of magnetization contour in (1.53 -1 0) plane which is indicated by green shading in (a); (c) Brillouin zone; (d)(e) spin degenerate energy bands (d) when SOC is off and (e) when SOC is on. The electronic properties are calculated by DFT method using PBE+U functional.

## B. Cubic, Fm-3m NiO illustrating no spin splitting SST-2

We use the experimental crystal structure from Ref. [38] for the undeformed rock-salt NiO and set the effective $U$ on Ni-3d orbits to 4.6 eV as the input for DFT electronic band structure calculations. Figure 3(a) shows the crystal structure of the AFM phase of the undeformed NiO. We apply the magnetic moment on $Ni^{2+}$ ($3d^8$) initially along the [11-2] direction, and obtain from our calculation a final magnetic moment on $Ni^{2+}$ ($3d^8$) of $1.7\mu_B$ along the [11-2] direction (see AFM magnetization contour in (110) plane in Fig. 3(b)) which agrees with the neutron-scattering measurements of 1.9 $\mu_B$ [39]. Figure 3(c) shows the first Brillouin zone and high symmetry k-paths. Figure 3(d) and (e) show the corresponding spin degenerate bands with SOC off and on. We see a direct gap at $L$ of 3.55 eV and a smaller indirect gap of 2.98 eV between VBM at L and CBM at some k point on Γ-K path. These values are smaller than the experimental 4.3 eV gap obtained from the combined photoemission/inverse photoemission measurement [40]. A better agreement on the band gap can be achieved via increasing the $U$ value, but this is outside the scope of the current paper. Only negligible difference on band structure near Fermi level has been predicted between (d) and (e), which is expected as both Ni and O are light elements. Such spin degenerate bands in the whole Brillouin zone even when SOC is on, are in agreement with our prediction of no spin splitting for SST-2.

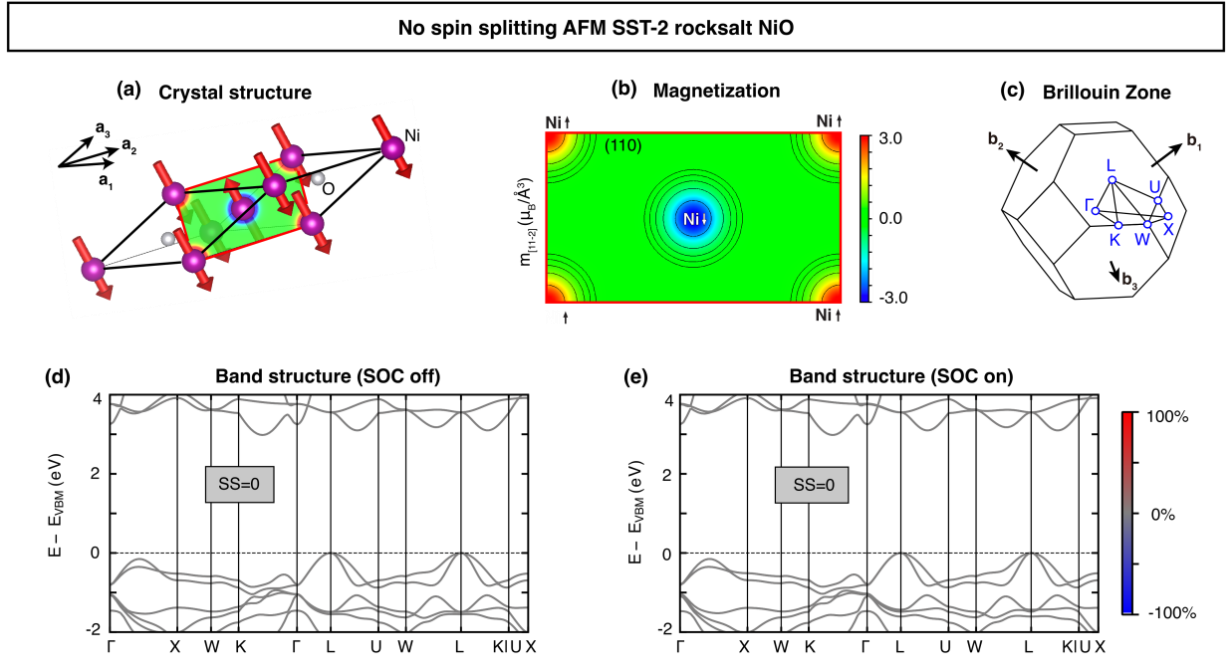

**Figure 3 | No spin splitting in AFM Rocksalt-NiO (AFM SST-2).** (a) Crystal structure and magnetic moments, where red arrows are used to indicate local magnetic moments; (b) the contour plot of magnetization along [11-2] direction in (110) plane which is indicated by green shading in (a); (c) the first Brillouin zone and high symmetry k-paths; (d)(e) spin-degenerate energy bands (d) when SOC is off and (e) when SOC is on. The electronic properties are calculated by DFT method using PBE+U functional.

## C. Orthorhombic, Pa-3 MnS$_2$ illustrating SOC- induced spin splitting SST-3A

We use the experimental crystal structure from Ref. [41] for the orthorhombic MnS$_2$ and set the effective $U$ on Mn-3d orbits to 5 eV as the input for DFT electronic band structure calculations. Figure 4(a) shows the crystal structure of the AFM MnS$_2$. We apply magnetic moment on Mn initially along the [001] direction, and obtain from the calculation a final magnetic moment on Mn f 4.6 $\mu_B$ aligned along the [001] direction (see AFM magnetization contour in (001) plane in Fig. A3(b)) which agrees with the electron configuration of S=5/2 for Mn$^{2+}$ in MnS$_2$. Figure 4(c) shows the Brillouin zone and high symmetry k-paths. Figure 4(d) and (e) show the corresponding spin degenerate bands with SOC off and on. We see an indirect gap of 2.98 eV between VBM at Γ and CBM at T. Only negligible difference on band structure near Fermi level has been predicted between (d) and (e), which is expected as both Mn and S are light elements. Such spin splitting present only when SOC is on agrees with our prediction for SST-3.

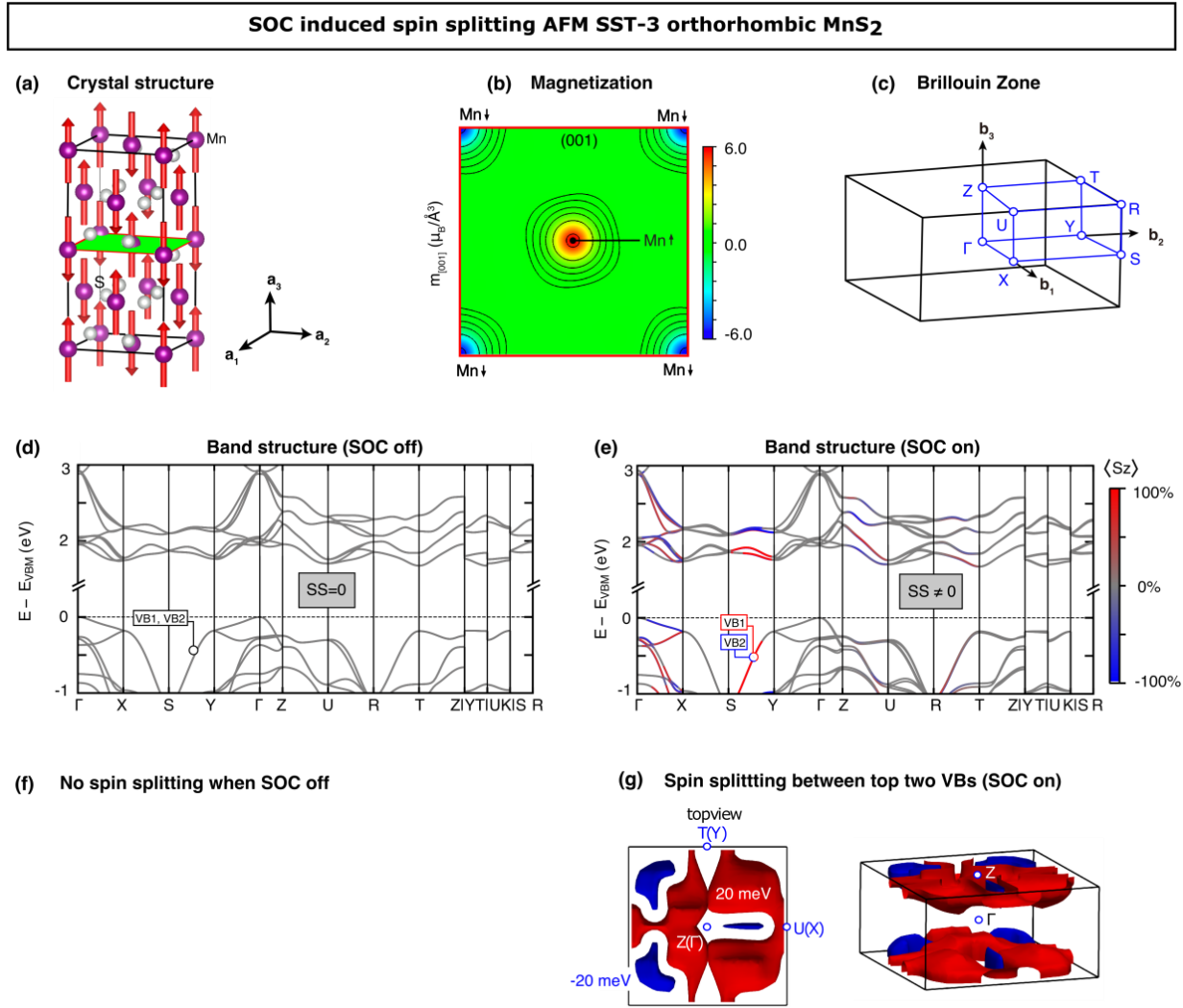

**Figure 4 | SOC induced spin splitting in AFM MnS$_2$ in orthorhombic unit cell (AFM SST-3A).** (a) Crystal structure and magnetic moments, where red arrows are used to indicate local magnetic moment; (b) magnetization contour plot in the middle (001) plane; (c) Brillouin zone and high symmetry k-paths. (d) Spin degenerate energy bands when SOC is off and (e) spin split energy bands when SOC is on. (f-g) Isosurface of spin splitting between top two valence bands (VB1, VB2) at 20 meV (red surface) and -20 meV (blue surface) in the Brillouin zone of top view and side perspective view when SOC is off (f) and when SOC is on (g); The electronic properties are calculated by DFT method using PBE+U functional.

## D.  Orthorhombic, Pnma LaMnO$_3$ illustrating collinear, centrosymmetric AFM-induced SST-4A

The Antiferromagnetic gapped insulator LaMnO$_3$ (Néel temperature T$_N$=139.5K [42]) has many desired electronic and magnetic properties induced by doping and pressure [43,44]. Historically, the insulating property of LaMnO$_3$ and similar 3d oxides are attributed to Mott-Hubbard interaction, but recent work showed that mean-field band theory (including the current work, see Fig. 5) could also correctly describes such compounds as insulators even Hubbard U=0 [45]. On heating through the Jahn-Teller temperature (T$_{JT}$=750K) the material becomes a ferromagnetic metal [46] and exhibits fully spin polarized conduction band [47] which makes the material an ideal candidate for spin electronic applications. Below the Néel temperature, the ground-state AFM LaMnO$_3$ has a centrosymmetric orthorhombic crystal structure of space group Pnma (also notated as Pbnm), and a magnetic structure of MSG Pn'ma'[48], which belongs to SST-4A. The single crystal sample of LaMnO$_3$ can be prepared using a floating method [48]. Moussa et. al. determined the ground-state crystal and magnetic structures via Neutron diffraction as shown in Figure 4(a): The local spins are on Mn$^{3+}$ ions, all collinearly pointing to $[010]$ and $[0\bar{1}0]$ directions; neighboring (001) planes hold opposite spin directions, i.e., the local spin is ferromagnetically coupled within the (001) plane and antiferromagnetically coupled between neighboring (001) planes (which is known as an AFM-A order) [48] [49].

We use experimentally observed ground-state crystal structure and magnetic moments[48] (Fig. 5(a)) and set the effective $U$ on Mn atoms to 3 eV as the input for DFT electronic band structure calculations. The local magnetic moment from our results is 3.77 $\mu_B$ which agrees well with other DFT predictions (see Table I of Ref. [50]). Fig. 5(b) shows the contour plot for magnetization of y-component on the (001) plane (green shading plane in Fig. 5(a)). Fig. 5(c) is the 3D view of the primitive Brillouin zone (and several high-symmetric k-points) of the Pn'ma' phase. Figure 5(d) and (e) show the calculated band structures of Pn'ma' LaMnO$_3$ on high symmetry k-paths of its Brillouin zone, with SOC (Fig. 5(d)) and without SOC (Fig. 5(e)) effect. Only negligible difference on band structure near Fermi level has been predicted between (d) and (e), which is expected as the states near Fermi level are from Mn and O, both of which are light elements. The bands show an indirect gap of approximately 1 eV between valence band maximum on Z-U near U and conduction band minimum at Γ. The top 2 (indexed by energy) valence bands (counting spin channels) have been denoted as VB1 and VB2 in Fig. 5(d) and (e). We see in Fig. 5(d) and (e) that the resulting band structures along high symmetry lines show only zero spin splitting (each band is evenly degenerated) and vanishing spin polarization (mapped by grey color) for all states. Indeed, for Pn'ma' LaMnO$_3$ we find the spin splitting and spin polarization occur even when SOC is absent, not on the high symmetric k-paths but at generic k-points away from these k-paths. Figure 5(f) and (g) show the 3D isosurfaces of the spin splitting in the first Brillouin zone between the top two valence bands (denoted as VB1 and VB2) for SOC off and SOC on cases, respectively. The spin splitting is evaluated by the eigenvalue of the spin-up state minus the eigenvalue of the spin-down state.

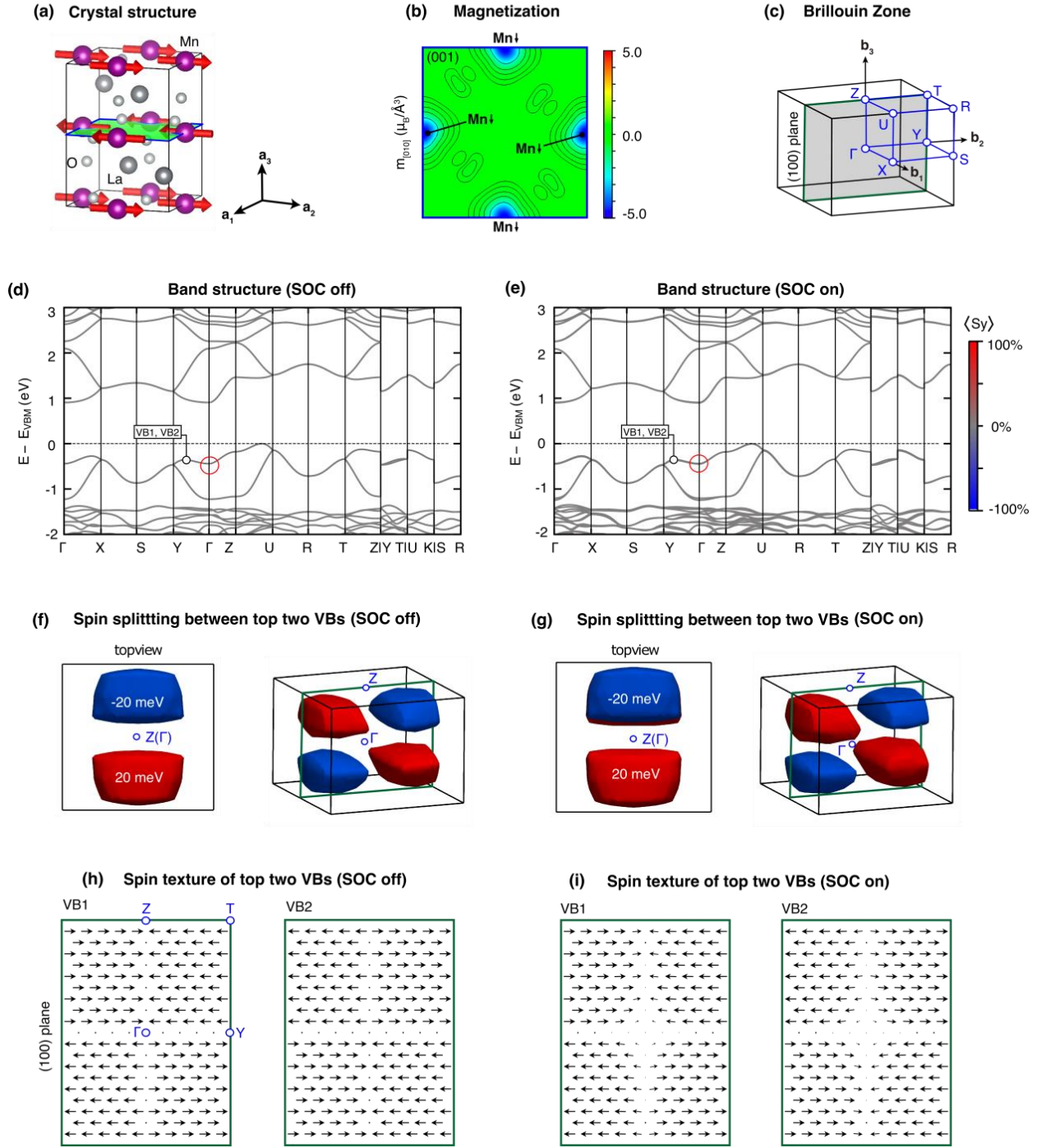

**Figure 5 | Spin polarization and spin splitting in centrosymmetric collinear AFM orthorhombic LaMnO₃ (AFM SST-4A).** (a) Crystal structure and magnetic moments, where red arrows are used to indicate local magnetic moments; (b) the contour plot of y-component of magnetization contour in (001) plane which is indicated by green shading in (a); (c) the first Brillouin zone; (d) energy bands when SOC is off and (e) energy bands when SOC is on. (f)(g) Isosurfaces of spin splitting in the first Brillouin zone between the top two valence bands (VB1, VB2) at 20 meV (red) and -20 meV (blue) when SOC is off (f) and when SOC is on (g). (i)(h) Spin textures of the top two valence bands VB1 and VB2 when SOC is off (h) and when SOC is on (i). The electronic properties are calculated by DFT method using PBE+U functional.

When the spin-up state is above the spin-down state the spin splitting takes a positive value (marked red in Fig. 5(f)(g)), and when the spin-up state is below the spin-down state the spin splitting takes a negative value (marked blue in Fig. 5(f)(g)). It can be seen that spin splitting (i) *exist in the Brillouin zone even when SOC if off, i.e., an AFM-induced spin splitting*, but (ii) *require a search over generic k-points instead of only high symmetry k-paths* (as the k-paths used in Fig. 5(d)(e)).

Figure 5(h) and (i) show the cross section (on the green rectangle plane in Fig. 5(g)) of the spin polarization in momentum space, for SOC off and SOC on cases. Here we show only VB1 and VB2, but we also consider the lowest conduction band as equal importance for potential applications. Because of the collinear spin along [010] direction, the AFM-induced spin polarization is also collinearly aligned along [010]. Similar to the case of AFM MnF₂[10], LaMnO₃ shows a 4-quadrant pattern where neighboring quadrants hold opposite spin polarizations. As we include SOC effect, the weak spin-orbit interaction of Mn and O does not change the 4-quadrant pattern, but induces a slight non-collinearity in the spin texture, especially around Γ point, where notable tilting of spin polarization away from [010] can be seen as shown in Figure 2(i). Such spin splitting and spin polarization present at generic k points even in the non-relativistic limit (i.e., when SOC is off) agree with our predictions for the SST-4A materials.

### E. Cubic Pa-3 NiS₂ illustrating non-collinear, centrosymmetric AFM-induced SST-4A

The narrow band gap semiconductor, non-collinear antiferromagnet NiS₂ (Néel temperature $T_N$=39.2K [51,52]) is an important model of Mott insulator and is reported to exhibit insulator-to-metal transition[42] by chemical substitution of S for Se[51]. Below the Neel temperature, the ground-state AFM NiS₂ has a centrosymmetric cubic pyrite crystal structure of space group Pa-3, and a non-collinear magnetic structure of MSG Pa-3[51] which belongs to SST-4A. The crystal consists of octahedral bonded NiS₆ connected in a face-centered-cubic of Ni sublattices. Katsuya et. al. measured the magnetic structure of single crystal NiS₂ using neutron diffraction method at 4.2K [51] as shown in Fig. 6(a): Magnetic structure of NiS₂ corresponds to a $\Gamma_1\psi_1$ representation with propagation vector $k = (0,0,0)$; specifically, the magnetic moments on Ni are antiferromagnetically arranged and aligned in the <111> directions.

# Non-collinear AFM-induced SST-4A cubic NiS$_2$

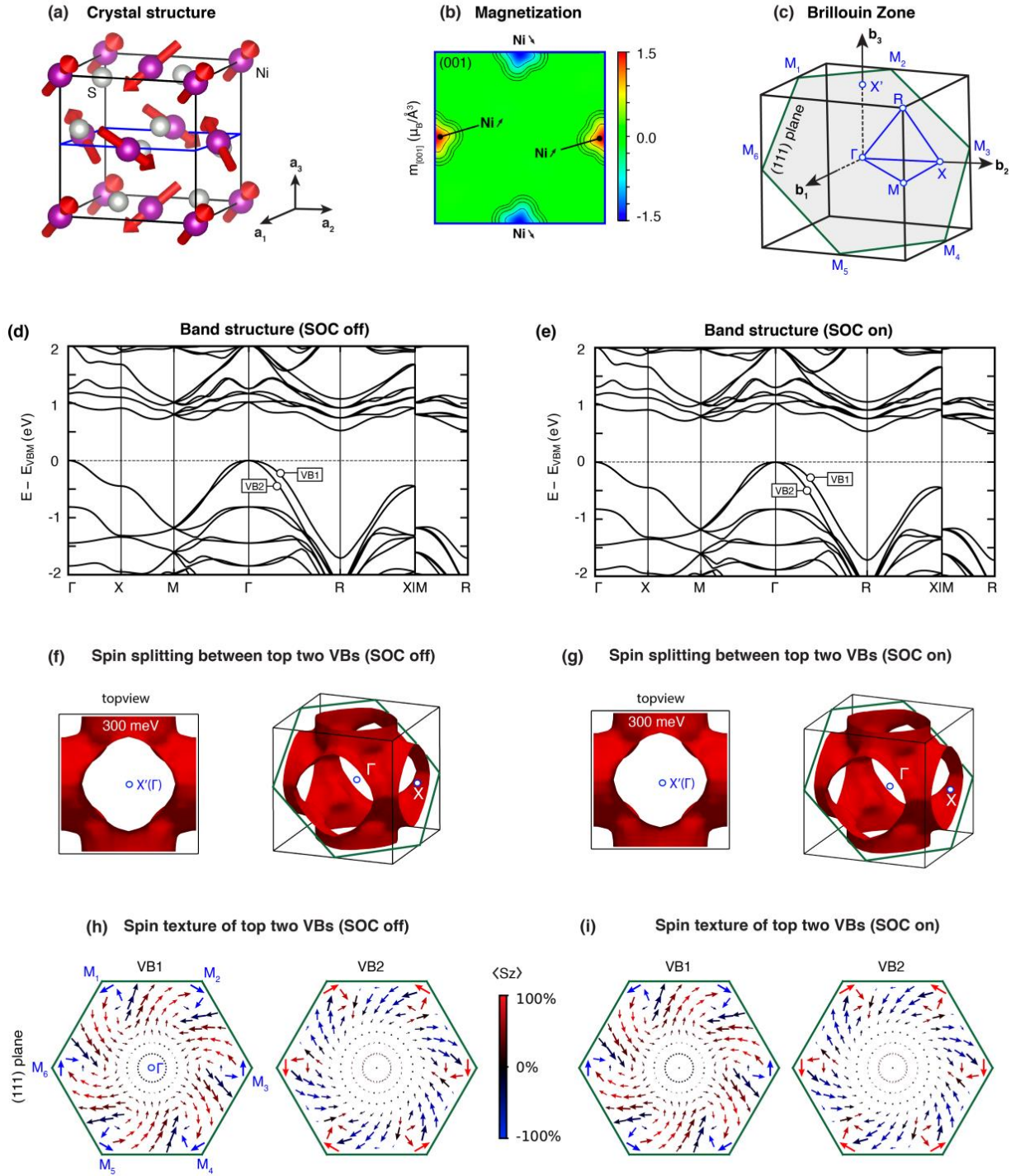

**Figure 6 | Spin polarization and spin splitting in centrosymmetric non-collinear AFM cubic NiS$_2$ (AFM SST-4A).** (a) Crystal structure and magnetic moments, where red arrows are used to indicate local magnetic moments; (b) the contour plot of z-component of magnetization in (001) plane which is indicated by blue rectangular in (a); (c) the first Brillouin zone; (d) energy bands when SOC is off and (e) energy bands when SOC is on. (f)(g) Isosurfaces of spin splitting between the top two valence bands (VB1, VB2) at 300 meV (absolute value) in the Brillouin zone when SOC is off (f) and when SOC is on (g). (i-h) Spin textures of the top two valence bands VB1 and VB2 when SOC is off (h) and when SOC is on (i). The electronic properties are calculated by DFT method using PBE+U functional.

We use the experimentally observed ground-state crystal and magnetic structures [53] (Fig. 6(a)) and set the effective *U* on Ni atoms to 4.6 eV as the input for DFT electronic band structure calculations. The calculated local magnetic moment is 1.4 $\mu_B$ which agrees well with the experimental results of 1.0 $\mu_B$ [53]. Fig. 6(b) shows the contour plot for magnetization (only z component) on one of the (001) plane (green shading square plane in Fig. 6(a)). Fig. 6(c) is the 3D view of the primitive Brillouin zone (and several high-symmetric k-points) of the Pa-3 phase. Figure 6(d) and (e) show the calculated band structures of Pa-3 NiS$_2$ on high symmetry k-paths of its Brillouin zone, with SOC (Fig. 6(d)) and without SOC (Fig. 6(e)) effect. Only negligible difference on band structure near Fermi level has been predicted between (d) and (e), because both Ni and S manifest very weak SOC strength. The bands show an indirect gap of approximately 0.5 eV between valence band maximum at Γ and conduction band minimum at R. The top 2 (indexed by energy) valence bands (counting spin channels) have been denoted as VB1 and VB2 in Fig. 6(d) and (e). Figure 6(f) and (g) shows the spin splitting isosurfaces of 300 meV between top two valence bands (VB1 and VB2) in the first Brillouin Zone. Because of the noncollinearity, it is not able to define spin up and spin down channels. Therefore, we evaluate the spin splitting only by the absolute value between two neighboring bands holding opposite spin polarization. The amplitude of spin splitting (i) increases as leaving the center Γ point, and (ii) is strongly anisotropic along different directions, e.g., the splitting vanishes (0 meV) along Γ-X, becomes non-zero (~100 meV) along Γ-M, and becomes significantly large (~300 meV) along Γ-R. (iii) SOC only has negligible effect on the spin splitting. Results (i)-(iii) agrees with the line-style band structures in Fig. 6(d) and (e).

The encouraging results lie in Figure 6(h) and (i), which show the cross section (on the green (111) plane in Fig. 6(g)) of the spin polarization in momentum space for SOC off and SOC on cases. Here we show only VB1 and VB2, but we also consider the lowest conduction band as equal importance for potential applications. The spin texture (i) is noncollinear inherited from its noncollinear magnetization; such noncollinear spin texture resembles those nontrivial spin texture (such as Rashba and Dresselhaus-type spin texture [4,5]) induced by relativistic SOC effect, but (ii) occurs even in the absence of SOC; (iii) the spin texture around the center point Γ show a 3-fold symmetry, and its pattern is different than the conventional Rashba or Dresselhaus pattern, indicating another mechanism (AFM magnetic mechanism).

Including SOC induces only negligible changes in spin texture comparing Figure 6(h) and (i). Such spin splitting and spin polarization present at generic k points even in the non-relativistic limit (i.e., when SOC is off) agree with our predictions on the SST-4A materials.

### F. Rhombohedral R3c MnTiO$_3$ illustrating collinear, non-centrosymmetric AFM-induced SST-4B

The high-pressure form MnTiO$_3$ which adopts the acentric LiNbO$_3$-type structure[54] is a multiferroic material. The multiferroic property of the compound makes it candidate material for potential applications in memory technologies [55,56]. The compound takes G-type antiferromagnetic order as well as ferroelectric orders. [54]. Besides, the compound also exhibits very weak ferromagnetism which has been neglected in the DFT modeling since such weak ferromagnetism has very small impact in the resulting spin splitting. Below the Néel temperature (T$_N$=28K), the magnetic moment on Mn align collinearly to (010) direction in the basal plane and align oppositely in the neighboring planes [54]. The AFM MnTiO$_3$ has non-centrosymmetric parent space group of R3c and magnetic space group of Cc' (no ΘI symmetry, MSG type III) with propagation vector $k = (0,0,0)$ which belongs to SST-4B.

In the calculation, we adopt the experimental structure from Ref. [54] and set the effective *U* on Mn atoms to 3 eV as the input for the DFT electronic band structure calculations. Figure 7(a) shows the crystal structure of AFM MnTiO$_3$. The calculated local magnetic moment is 4.5 $\mu_B$ which is close to the experimental results of 3.9 $\mu_B$[54]. Fig. 7(b) shows the contour plot for y-component of magnetization on one of the (012) plane (blue outlined square plane in Fig. 7(a)). Figure 7(c) is the 3D view of the primitive Brillouin zone (and several high-symmetric k-points) of the R3c phase. Figure 7(d) and (e) show the calculated band structures of MnTiO$_3$ on high symmetry k-paths of its Brillouin zone, with SOC (Fig. 7(d)) and without SOC (Fig. 7(e)) effect. Negligible difference on band structure near Fermi level has been found between (d) and (e), because Mn, Ti and O manifest very weak SOC strength. The bands show an indirect gap of 1.9 eV when SOC is off which is larger than 0.85 eV reported in Ref. [57]. The top 2 (indexed by energy) valence bands (counting spin channels) have been denoted as VB1 and VB2 in Fig. 7(d) and (e). Figure 7(f) and (g) show the spin splitting isosurfaces of 45 meV (red surface) and -45 meV (blue surface) between the top two valence bands (VB1 and VB2) in the first Brillouin zone. It can be seen that the spin splitting (i) *exist in the Brillouin zone even when SOC if off, i.e., an AFM-induced spin splitting*, but (ii) *require a search over generic k-points instead of only high symmetry k-paths* (as the k-paths used in Fig. 7(d)(e)).

# Collinear AFM-induced SST-4B rhombohedral MnTiO₃

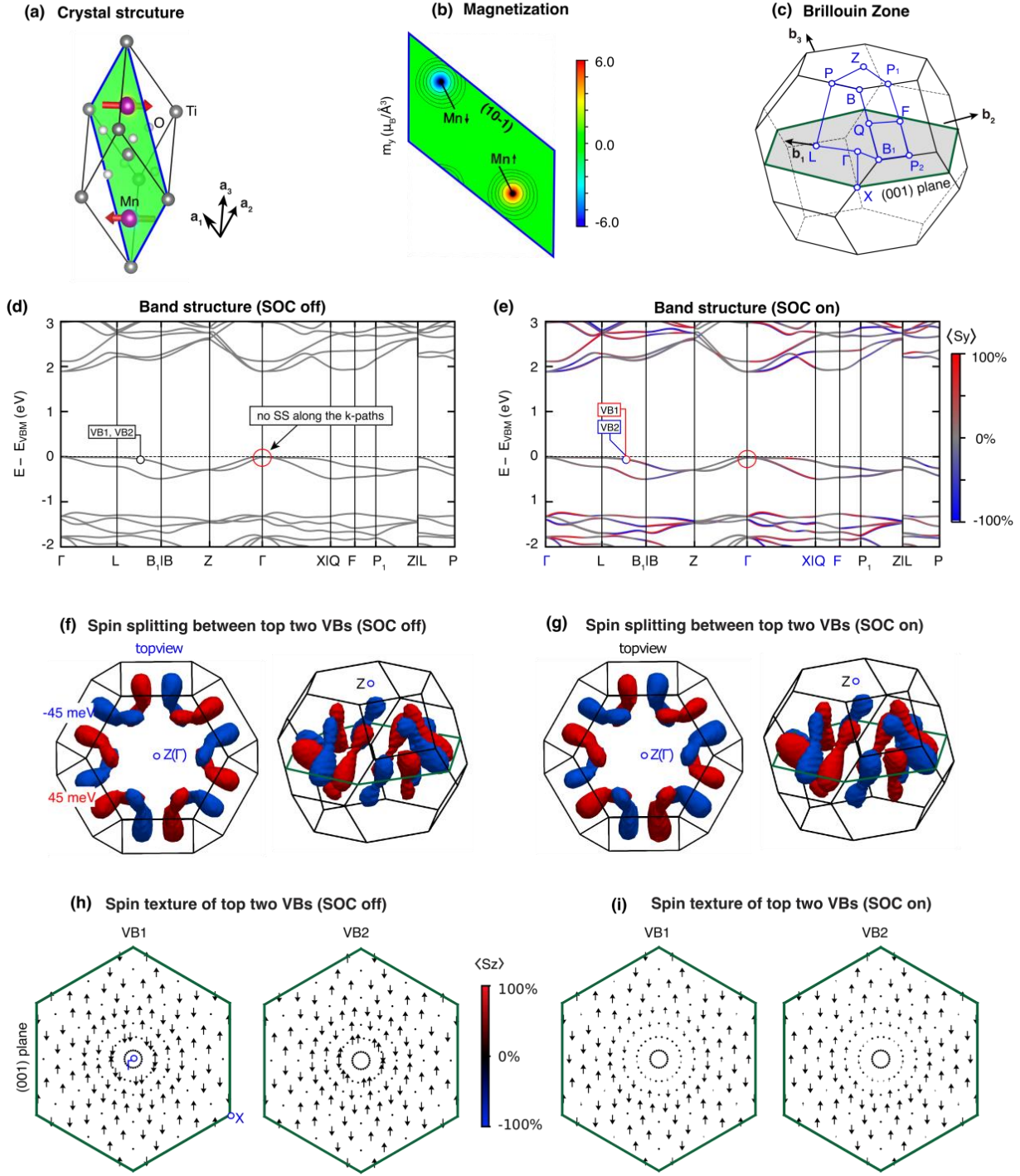

**Figure 7 | Spin polarization and spin splitting in non-centrosymmetric collinear AFM rhombohedral MnTiO₃ (AFM SST-4B).** (a) Crystal structure and magnetic moments; red arrows are used to indicate local magnetic moments; (b) the y- component of magnetization contour in (001) plane which is indicated by green shading in (a); (c) the first Brillouin zone; (d) energy bands when SOC is off and (e) energy bands when SOC is on. (f)(g) Isosurfaces of spin

splitting between the top two valence bands (VB1, VB2) at 45 meV (red) and -45 meV (blue) in the Brillouin zone when SOC is off (f) and when SOC is on (g). (i)(h) Spin textures of the top two valences bands VB1 and VB2 when SOC is off (h) and when SOC is on (i). The electronic properties are calculated by DFT method using PBE+U functional.

Figure 7(h) and (i) show the cross section (on the green (001) plane in Fig. 7(g)) of the spin polarization in momentum space for SOC off and SOC on cases. The spin polarization (i) collinearly aligns in the same direction as the collinear magnetic moment but (ii) varies in magnitude; (iii) the spin texture shows small noncollinearity when SOC is included. Such spin splitting and spin polarization are present at generic k points even in the non-relativistic limit (i.e., when SOC is off) which agrees with our prediction for SST-4B.

### G. Hexagonal P6$_3$cm ScMnO$_3$ illustrating non-collinear, non-centrosymmetric AFM induced SST-4B

AFM ScMnO$_3$ oxide and similar compounds of rare-earth manganites RMnO$_3$ (R=Sc, Y, Ho, Er, Tm, Yb) which stable in hexagonal crystal are a type of important multiferroic compounds[55,56]. Like MnTiO$_3$, because of the underlining multiferroic they are good candidate materials for nonvolatile memory devices [55,56]. Weak ferromagnetism is also found in ScMnO$_3$ which has been neglected in the DFT modeling since such weak ferromagnetism has very small impact in the resulting spin splitting. Below the Néel temperature (T$_N$=129K), the AFM ScMnO$_3$ has non-centrosymmetric parent space group of P6$_3$cm and magnetic space group of P6$_3$c'm' (no ΘI symmetry, MSG type III) with propagation vector $k = (0,0,0)$ which also belongs to SST-4B. Polycrystalline samples of hexagonal ScMnO$_3$ can be prepared by solid state reaction.[58] Neutron scattering experiments [59-61] show the magnetic moments contained in the (001) plane and oriented in the [100] direction.

In the calculation, we adopt the experimental structure from Ref. [54] and set the effective *U* on Mn atoms to 2.5 eV as the input for the DFT electronic band structure calculations. Figure 8(a) shows the crystal structure of AFM ScMnO$_3$. The calculated local magnetic moment is 4.0 $\mu_B$ which is close to the experimental results of 3.5 $\mu_B$ measured by neutron diffraction at 1.7 K[59]. Figure 8(b) shows the contour plot for magnetization of x-component on one of the (001) plane (blue outlined diamond plane in Fig. 8(a)). Figure 8(c) is the 3D view of the primitive Brillouin zone (and several high-symmetric k-points) of the P6$_3$cm phase. Figure 8(d) and (e) show the calculated band structures of ScMnO$_3$ on high symmetry k-paths of its Brillouin zone (shown in Figure 8(c)). The bands show a quasi-direct gap of approximately 0.8 eV at Γ when SOC is off which is smaller to the calculated value of 2.5 eV reported in Ref. [62]. Negligible difference on band structure near Fermi level has been found between (d) and (e), because Mn, Sc and O manifest very weak SOC strength.

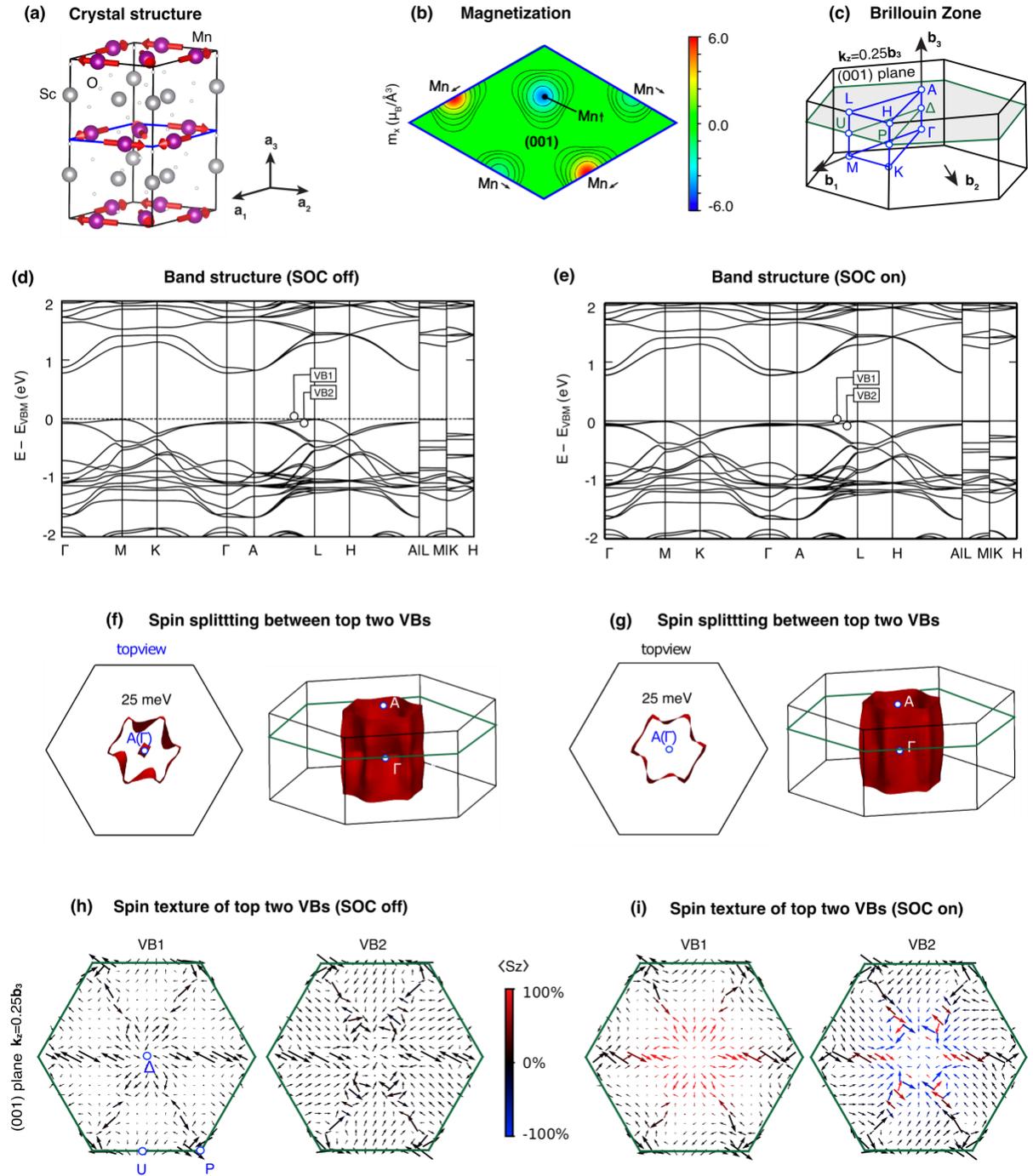

**Figure 8 | Spin polarization and spin splitting in non-centrosymmetric non-collinear AFM hexagonal ScMnO₃ (AFM SST-4B).** (a) Crystal structure and magnetic moments; red arrows are used to indicate local magnetic moments; (b) x-component of magnetization contour plot in (001) plane which is indicated by green shading in (a); (c) Brillouin zone; (d) energy bands when SOC is off and (e) energy bands when SOC is on. (f-g) Isosurfaces of spin splitting between the top two valence bands (VB1, VB2) at 25 meV (absolute value) in the BZ of top view and side perspective

when SOC is off (f) and when SOC is on (g). (i-h) Spin textures of the top two valence bands VB1 and VB2 when SOC is off (h) and when SOC is on (i). The electronic properties are calculated by DFT method using PBE+U functional.

The top 2 (indexed by energy) valence bands (counting spin channels) have been denoted as VB1 and VB2 in Fig. 8(d) and (e). Figure 8(f) and (g) show the spin splitting isosurfaces of 25 meV (red surface) between the top two valence bands (VB1 and VB2) in the first Brillouin zone. It can be seen that the spin splitting (i) *exist in the Brillouin zone even when SOC if off, i.e., an AFM-induced spin splitting*, and (ii) appears on $k_x$ and $k_y$ directions (Γ-M-K, A-L-H) but vanish on $k_z$ direction (Γ-A, K-H, M-L).

Figure 8(h) and (i) show the cross section (selected (001) plane with $k_z = \frac{1}{4}(2\pi/c)$) of the spin polarization in momentum space for SOC off and SOC on cases. Again, because of the noncollinearity magnetization, the spin texture is (i) noncollinear – various in orientation in momentum space, and (ii) tilts the slightly out of plane when SOC is included as represented by colored arrows (especially around the center Δ point) in Figure 8(i). The spin splitting and spin polarization are present at generic k points even in the non-relativistic limit (i.e., when SOC is off) which agrees with our prediction for SST-4B.

## V. Discussion of utility of AFM vs FM spintronics and collinear vs noncollinear AFM spin textures

The generation of active spin polarization has traditionally been based on non-zero net magnetization and spin-orbit coupling. Current technology of spintronic are based on ferromagnets (mostly collinear) [1,3]. Antiferromagnets, on the other hand, have alternate local magnetic moments on different atomic sites that mutually compensate, leading to a global zero net magnetization. They are thus unresponsive to external magnetic field and have been considered for a long time useless for field effect applications, [63] but restricted to passive role as exchange-bias materials [64]. Yet, there are certain possible advantages of AFM over FM for applications: (i) AFM compounds are more abundant than FM compounds and often have higher transition temperatures. [65] (ii) AFM systems generally have faster dynamic than FM systems [66,67]; (iii) AFM systems are insensitive to magnetic perturbation and do not suffer from stray field[66,67]. These features point to a more power efficient, smaller, faster operating, and robust AFM based spintronic scenario. The newly discovered spin polarized electron states in AFM [10] could promote the previously dismissed materials of AFM to an equal footing as FM materials for spin electronic.

Advantages and disadvantages of different spin polarization mechanism (FM, AFM, SOC), and collinear and noncollinear have been summarized in Table I.

**Table I | Advantages and disadvantages of collinear FM, collinear AFM, noncollinear magnets and SOC materials in terms of physical effects and potential applications.**

| Systems | Properties | Has | Does not have |
|---|---|---|---|
| **Collinear FM (non-relativistic)** | Collinear spin texture; k-dependent spin splitting | (1) Spin conservation; (2) Spin polarized currents; Spin transport torque (STT) | (1) Current-driven magnetization (i.e., no current induced torque by the material on itself); (2) Dissipationless charge or spin current induced by electric field; (i.e. no spin Hall effect, no anomalous hall effect) |
| **Collinear AFM (non-relativistic)** | Collinear spin texture; k-dependent spin splitting | (1) Spin conservation; (2) Magnetic spin Hall effect[a]; spin polarized currents[a]; Spin transport torque (3) Ultrafast dynamics (4) Absence of magnetic stray fields (5) Robust against external magnetic field | (1) Current-driven magnetization (i.e., current induced torque by the material on itself); (2) Dissipationless charge or spin current induced by electric field; (i.e. no spin Hall effect, no anomalous hall effect) |
| **Noncollinear AFM (non-relativistic)** | Noncollinear spin texture; k-dependent spin splitting | (1) Spin Hall effect[b]; (2) Anomalous Hall effect[c]; (3) Spin polarized current[d]; (4) Current-driven magnetization | Spin conservation, long spin lifetime; |
| **SOC Rashba and Dresselhaus effect (relativistic)** | Noncollinear spin texture; k-dependent spin splitting | (1) Spin Hall effect[b]; (2) Anomalous hall effect[c]; (3) Spin polarized current[d]; (4) Current-driven magnetization (SOT); | Spin conservation, long spin lifetime; |

a. Ref. [68]
b. Ref. [28]
c. Ref. [69]
d. Ref. [27]

## VI. Conclusions

In this work we have studied the AFM-induced spin splitting and spin polarization effects. Such effects could exist even in the absence of SOC and even in centrosymmetric structures, in both collinear and noncollinear antiferromagnets. Starting from the symmetry design principles that enable such AFM-induced spin splitting effect, we can generally divide materials of different symmetry into seven spin splitting prototypes. We classify all the 1651 3D magnetic space groups based on the design principles into these 7 different categories so that one can predict the spin splitting and spin polarization behavior of one compound given its magnetic space group. We further apply the symmetry rules to examine a board set of known antiferromagnetic materials included in the Bilbao MAGNDATA database[31]. We find 422 magnetic space groups and a list of magnets that can hold the AFM-induced, SOC-independent spin splitting and spin polarization. We examine the band structures, spin splitting and spin texture of specific subset of these including both collinear and noncollinear AFM. We find noncollinear spin texture in noncollinear AFM that resembles SOC induced momentum-dependent spin polarization. This work then provides the foundation of AFM spin polarization, offering also a bridge between such design principles and real-life crystals and magnetic structures.

## ACKNOWLEGEMNETS

We thank Emmanuel Rashba for many discussions that led to Ref. [10] and much more. The work at CU Boulder was supported by the National Science Foundation (NSF) DMR-CMMT Grant No. DMR-1724791 that supported the formal theory development of this work. The electronic structure calculations of this work were supported by the U.S. Department of Energy, Office of Science, Basic Energy Sciences, Materials Sciences and Engineering Division under Grant No. DE-SC0010467. J.-W.L. was supported by the National Natural Science Foundation of China (NSFC) under Grant No. 61888102. This work used resources of the National Energy Research Scientific Computing Center, which is supported by the Office of Science of the U.S. Department of Energy.